\documentclass{article}
\usepackage{amsmath}
\usepackage{amsfonts}
\usepackage{amssymb}
\usepackage{graphicx}%
\begin{document}
\markboth{Akihisa Tomita}
{Measured Quantum Fourier Transform}

\title{Measured Quantum Fourier Transform of 1024 Qubits on Fiber Optics
}

\author{Akihisa Tomita \\
Quantum Computation and Information Project, ERATO, JST\\
Miyukigaoka 34, Tsukuba, Ibaraki 305-8501, Japan\\tomita@qci.jst.go.jp \and Kazuo Nakamura \\
Fundamental Research Laboratories, NEC Corporation\\
Miyukigaoka 34, Tsukuba, Ibaraki 305-8501, Japan\\nakamura@frl.cl.nec.co.jp}
\maketitle

\begin{abstract}
Quantum Fourier transform (QFT) is a key function to realize quantum
computers. A QFT followed by measurement was demonstrated on a simple
circuit based on fiber-optics. The QFT was shown to be robust against
imperfections in the rotation gate. Error probability was estimated to be 0.01
per qubit, which corresponded to error-free operation on 100 qubits. The error
probability can be further reduced by taking the majority of the accumulated
results. 
The reduction of error probability resulted in a successful QFT demonstration on
1024 qubits.
\end{abstract}

\section{Introduction}

Quantum computers have been expected to posses more computational ability over
conventional computers. Shor's factorization algorithm\cite{Shor97}, for
example, has proved the power of quantum computation, which resolves a
composite number $N$ ($L$ bits long) into the prime factors in the polynomial
time of $L$, whereas all the known classical algorithms require exponential
resources. The factorization problem is also important from a practical point
of view, because efficient algorithms can break the RSA public-key cryptosystem.

Although Shor's algorithm and its related algorithms are well established in theory,
practical implementations of them are still in their infancy. So far, only an
NMR-based computer has demonstrated the factorization of 15 ($=3\times5$) with
seven qubits\cite{Vandersypen01}. Until now, it has been difficult to see how
one could increase the number of qubits to solve problems of practical
significance, such as breaking the RSA public-key cryptosystem. One of the
main obstacles to realize large scale quantum computers is errors in gate operations
due to decoherence. Theory of fault-tolerant quantum
computation\cite{Preskill98a} shows that arbitrary-length error-free
computation can proceed with only a polynomial overhead in time and space, so
long as the error probability per fundamental operation is kept below a
threshold value. Typical predicted values are in 10$^{-4}$ to 10$^{-6}$, which
depend on the product of qubits and gate operations required in the
computation, and on the model of decoherence\cite{KLZ98,Preskill98b}. These
threshold values are still too low for the present technology, so that it is
desirable to reduce the requirement in the scale of quantum computation.

Recently, Beauregard\cite{Beauregard03} showed that $2L+3$ qubits are enough
to perform Shor's algorithm. He used quantum Fourier transform (QFT) for
modular exponentiation, and a QFT followed by measurement (we will refer it to
MQFT after Measured Quantum Fourier Transform) on the control
qubit\cite{GN96,PP00}. It would be remarkable that MQFT is
fault-tolerant\cite{GN96}, so that the above scheme not only reduces the

number of qubits but also provides robustness against decoherence.

One of the main issues in implementing quantum
information technology is selection of quanta to carry the information.
Among a number of the implementation proposals, photons have played an
important role from the beginning of the practical proposal on quantum information technology.
Photons have advantages to implement qubits: SU(2) space easily implemented by polarization, very
weak coupling to the environment, and existing single photon measurement technique. 
Moreover, we can utilize fruit of extensive research and development efforts of the optical communication
industry. Commercially available fiber-optic devices enable us to construct efficient quantum circuits that
consist of one-qubit operations (including classically controlled gates) with a reasonable cost.
Fiber-optics resolves the
mode matching problems in conventional optics, and provides mechanically stable optical circuits. 
It would be worth exploring feasibility of the quantum information processing based on fiber-optics.
In this article, we show an implementation of the MQFT on a circuit built up
with commercially available fiber-optic devices. We have achieved error-free
MQFT operation on 1024 qubits, despite the imperfections in the actual devices.
This shows that photonic devices are suitable to process quantum information.

In the rest of the article, we will report the MQFT experiment in detail.
We will review the QFT and MQFT used in the phase estimation algorithm in section 2.
The implementation of MQFT with fiber-optics devices will be given in section 3. We will also show 
successful result on MQFT over 255 qubits and 1024 qubits. 
In section 4, we will consider effects of the imperfections in the experimental apparatus, and show
the validity of the majority voting by repeated measurements. We will also discuss a direction toward
making quantum computers expanding the present MQFT implementation.

\section{Theory}

\begin{figure} [pts]
\begin{center}
\includegraphics[
height=4.7539in,
width=3.3295in
]%
{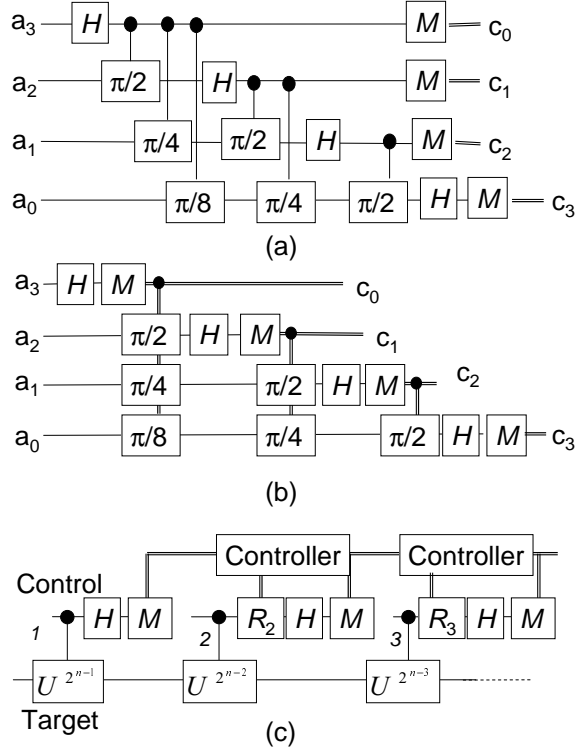}%
\caption {Quantum circuits for measured quantum Fourier transform. %
(a):A quantum circuit with controlled-rotation gates. %
(b):A quantum circuit with classically controlled-rotation gates\cite{GN96}. %
(c):A serial quantum circuit of phase estimation\cite{PP00}. %
 \textit{H} and \textit{M} stand for Hadamard gate and measurement, respectively. %
Phase of the controlled-rotation gates are represented by the numbers $\pi/2$, $\pi/4$, %
and $\pi/8$. $R_{k}$ is defined by (\ref{rotation}) in the text.}
\label{figure1}%
\end{center}
\end{figure}

The heart of the factorization and related quantum algorithms lies in the
phase estimation algorithm\cite{CEMM98} that consists of controlled-unitary
operations (c-$U$'s) and MQFT on the control qubits. The phase estimation problem
is given as follows: An eigenvalue of a
unitary transform $U$ defines a phase $\varphi$ as $U\left\vert u\right\rangle
=\exp\left[  2\pi i\,\varphi\right]  \left\vert u\right\rangle $. Our task is
to determine the phase expressed in $n$ bits by $\varphi=\varphi_{1}%
2^{-1}+\cdots+\varphi_{n}2^{-n}=0.\varphi_{1}\cdots\varphi_{n}$. 
When the target qubit state
is provided by the eigenstate $\left\vert u\right\rangle$ of the unitary transform $U$, 
the states after controlled-$U^{j}$ according to the superposition of 
all the possible control qubit state $\left\vert j\right\rangle$ is given by
\begin{equation}
\frac{1}{2^{n/2}}\sum_{j=0}^{2^{n}-1}\left\vert j\right\rangle%
U^{j}\left\vert u\right\rangle =\frac{1}{2^{n/2}}\sum%
_{j=0}^{2^{n}-1}e^{2\pi i\varphi j}\left\vert j\right\rangle%
\left\vert u\right\rangle .\label{cUs}%
\end{equation}
The inverse Fourier transform performs the transform
\begin{equation}
\frac{1}{2^{n/2}}\sum_{j=0}^{2^{n}-1}e^{2\pi i\varphi j}\left\vert
j\right\rangle \left\vert u\right\rangle \longrightarrow \left\vert \tilde{%
\varphi}\right\rangle \left\vert u\right\rangle ,\label{IFT}%
\end{equation}
where $\left\vert \tilde{\varphi}\right\rangle$ denotes a state which provide a good 
estimation of $\varphi$ when we measure the control qubits. 
In short, the controlled-$U^{j}$
provides an unknown phase according to the problem. MQFT then determines the
phase to find the solution. Though a text book\cite{NS00} shows QFT
implemented by a quantum circuit of Hadamard gates and controlled rotation
gates as in Fig. \ref{figure1}(a), an alternative form of the MQFT circuit was constructed
with only one-qubit gates as in Fig. \ref{figure1}(b)\cite{GN96}. This is based on the
fact that c-$U$'s commute with measurements when the control qubits are
measured in the computational basis. We can thus replace the
controlled-unitary gates with unitary gates controlled classically by the
results of the measurements\cite{NS00}. Since the latter devices act on only
the target qubit, they are much easier to realize than two-qubit gates. Though
Griffiths and Niu described their circuit as semiclassical\cite{GN96}, it is
worth to point out that the circuit is fully equivalent to the quantum circuit
constructed with controlled-rotation gates.

Parker and Plenio\cite{PP00} found that the phase estimation using MQFT can be
operated qubit by qubit with only one rotation at a time, as shown in Fig. \ref{figure1}(c). 
This reflects the fact that the RHS of (\ref{cUs}) can be rewritten by a product form as
\begin{equation}
\frac{1}{2^{n/2}}\left(  \left\vert 0\right\rangle +e^{2\pi i\,2^{n-1} \varphi}
\left\vert 1\right\rangle \right)  \left(  \left\vert 0\right\rangle +e^{2\pi
i\,2^{n-2} \varphi} \left\vert 1\right\rangle \right)  \cdots\left(
\left\vert 0\right\rangle +e^{2\pi i\,\varphi%
}\left\vert 1\right\rangle \right) \left\vert u\right\rangle  .\label{product-state}%
\end{equation}
The circuit operates on the target qubits, which are in an eigenstate of $U$, with
one control qubit at each step. In some applications, however, it is not easy to
find the eigenstates (if we know an eigenstate, the problem is already solved.)
Then, instead of using an eigenstate, we can start with a
superposition of the eigenstates:
\begin{equation}
\left\vert \psi_{0}\right\rangle =\frac{1}{\sqrt{r}}\sum_{s=0}^{r-1}%
c_{s}\left\vert u_{s}\right\rangle .\label{target0}%
\end{equation}
We observe $\left\vert 1\right\rangle =r^{-1/2}\sum\left\vert u_{s}%
\right\rangle $ in order-finding algorithm\cite{CEMM98}. The phase estimation
algorithm proceeds qubit by qubit as follows. In the first step, the c-$U^{2^{n-1}}$ with
respect to the first control qubit $2^{-1/2}\left(  \left\vert 0\right\rangle
+\left\vert 1\right\rangle \right)  $ transforms the initial state to an
entangled state%
\begin{equation}
U^{2^{n-1}}\left(  \left\vert 0\right\rangle +\left\vert 1\right\rangle
\right)  \left\vert \psi_{0}\right\rangle =\sum_{s=0}^{r-1}\left(  \left\vert
0\right\rangle +\exp\left[  \pi i\varphi_{n}^{s}\right]  \left\vert
1\right\rangle \right)  c_{s}\left\vert u_{s}\right\rangle ,\label{CU1}%
\end{equation}
where we use $\exp\left[ 2\pi i\,2^{n-1} \varphi\right]=%
\exp\left[ \pi i\, \varphi_{n}\right]$.
Here (and hereafter,) normalization constants are omitted for simplicity. Then
a Hadamard gate transforms the control qubit to $\left\vert 0\right\rangle $
or $\left\vert 1\right\rangle $ according to the bit value $\varphi_{n}%
^{s}=\left\{  0,1\right\}  $, and the total state is given by%
\begin{equation}
\left\vert 0\right\rangle \sum_{s\in\left\{  \varphi_{n}^{s}=0\right\}  }%
c_{s}\left\vert u_{s}\right\rangle +\left\vert 1\right\rangle \sum
_{s\in\left\{  \varphi_{n}^{s}=1\right\}  }c_{s}\left\vert u_{s}\right\rangle
.\label{H1}%
\end{equation}
The target qubit states collapse to either of the superposition $\sum
_{s\in\left\{  \varphi_{n}^{s}=0\right\}  }c_{s}\left\vert u_{s}\right\rangle
$ or $\sum_{s\in\left\{  \varphi_{n}^{s}=1\right\}  }c_{s}\left\vert
u_{s}\right\rangle $ by the measurement on the control qubit. The former
consists of the eigenstates belonging to the eigenvalues $0.x_{1}\cdots
x_{n-1}0$, and the latter consists of those belonging to the eigenvalues
$0.x_{1}\cdots x_{n-1}1$, where the bit values $x_{1}\cdots x_{n-1}$ are
arbitrary at this step. In the $k$ th
step, c-$U^{2^{n-k-1}}$ provides the state%
\begin{equation}
\sum_{s\in\left\{  \varphi%
_{n-k+2}^{s}\cdots\varphi_{n}^{s}=\varphi_{n-k+2}\cdots\varphi_{n}\right\}
}\left(  \left\vert 0\right\rangle +\exp\left[ 2\pi i\,0.\varphi_{n-k+1}^{s}%
\varphi_{n-k+2}\cdots\varphi_{n}\right] \left\vert 1\right\rangle \right)
c_{s}\left\vert u_{s}\right\rangle \label{CUk}%
\end{equation}
The control qubit state is reduced to $2^{-1/2}\left(  \left\vert
0\right\rangle +\exp[2\pi i\,0.\varphi_{n-k+1}^{s}]\left\vert 1\right\rangle
\right)  $ by applying the classically controlled rotation $R_{k}$ defined by%
\begin{equation}
R_{k}=%
\begin{pmatrix}
1 & 0\\
0 & \exp\left[  -2\pi i\Phi_{k}\right]
\end{pmatrix}
,\label{rotation}%
\end{equation}
where the rotation angle $\Phi_{k}$ is determined from the bit values obtained
in the previous measurements,\textit{ i.e.},%
\begin{equation}
\Phi_{k}=\sum_{j=1}^{k-1}\frac{1}{2^{j+1}}\varphi_{n-k+j+1}.\label{angle}%
\end{equation}
Measurement of the control qubit in the computational basis after Hadamard
transform again yields a deterministic value of $\varphi_{n-k+1}^{s}%
=\varphi_{n-k+1}$. Continuing the operation steps bit by bit
will provide all the bit values of $\varphi$ with probability proportional to
$\left\vert c_{s}\right\vert ^{2}$. The target qubit states collapse to one of
the binary branches according to the measurement result, and thus the number
of possible target qubit states is reduced at each step.

\section{Experiment}

\begin{figure} [pts]
\begin{center}
\includegraphics[
height=4.7539in,
width=3.3295in
]%
{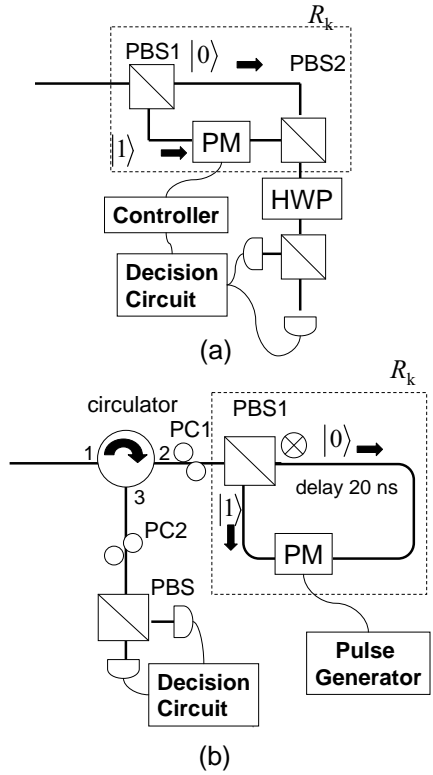}%
\caption{Quantum circuit for measured quantum Fourier transform implemented by fiber-optics. 
(a):the principle of the circuit. (b):practical circuit. PBS, PM, HWP, 
and PC stand for polarization beam splitters, phase modulator, half-wave plate,
and polarization controllers, respectively.}
\label{figure2}%
\end{center}
\end{figure}

Fig. \ref{figure2}(a) shows an implementation of MQFT. Qubits are represented by the
single photon polarization states, where the $\left\vert 0\right\rangle $ and
$\left\vert 1\right\rangle $ states are defined to be horizontally polarized
and vertically polarized, respectively. A sequence of single photon pulses
enters the quantum circuit. The input photons are generally elliptically
polarized according to the bit values of the phase $\varphi_{1},\cdots,\varphi_{n}$
 between the basis states
($\left\vert 0\right\rangle $ and $\left\vert 1\right\rangle $) to simulate
the output of the c-$U$'s as shown in (\ref{product-state}).

The key device of the MQFT circuit is the rotation gate, which is implemented
as an interferometer constructed with polarization beam splitters (PBS) and a
phase modulator (PM). The PBS1 divides the states according to the polarizations, and the PM provides a
relative phase shift given by (\ref{rotation}) to the $\left\vert
1\right\rangle $ state.  The two polarizations are combined again by the PBS2, and the
polarization state is analyzed by a half wave plate (HWP) and a PBS. The HWP
provides phase difference between the polarization. When its axis is tilted
22.5 deg, HWP transforms $\left\vert 0\right\rangle \rightarrow \left( \left\vert 0%
\right\rangle + \left\vert 1\right\rangle \right) / \sqrt{2}$ and 
$\left\vert 1\right\rangle \rightarrow \left( \left\vert 0%
\right\rangle - \left\vert 1\right\rangle \right) / \sqrt{2}$, thus
acts as a Hadamard gate.  
The accuracy of the phase shift is determined by the
precision of the electric pulse applied to the PM. In most pulse
generators, the precision is limited to three digits, \textit{i.e.}, 8-10
bits. This implies the rotation angle $\Phi_{k}$ in (\ref{angle}) should
be truncated at the $m$ th bit ($m<10$).

The interferometer in Fig. \ref{figure2}(a) will work in principle, however, 
is sensitive to disturbance in practice. We
employed a fiber loop configuration often referred to a Sagnac interferometer
as shown in Fig. \ref{figure2}(b), where orthogonally polarized photons propagate in
opposite directions through the same fiber. The two basis states are subject
to the same additional phase fluctuation, so that the present MQFT circuit is
robust to disturbances. The fiber-loop rotation gate consisted of a
circulator, a polarization controller (PC), a PBS, a PM, and polarization
maintaining fiber (PMF). The circulator is a unidirectional device that allows
light propagation only in the $1\rightarrow2$ and $2\rightarrow3$ direction.
Before entering the PBS, photons were passed through PC1 to compensate for the
unintentional polarization rotation due to the birefringence in the fiber. The
photons were divided by PBS1. The output ports of the PBS1 were both aligned
to the slow axis of the PMF. The horizontally polarized $\left\vert
0\right\rangle $ state propagated in the clockwise (CW) direction and the
vertically polarized $\left\vert 1\right\rangle $ state in the counter
clockwise (CCW) direction. Note that the PBS converts the polarizations 
into the directions of the propagation,
and that the qubits were represented by CW and CCW pulses 
with the same polarization in the fiber-loop.
 
The PM was placed at an asymmetric position in the
fiber loop to provide the relative phase. 
The CW pulses entered the PM earlier than the CCW pulses for 20 ns
in the present setting. The arrival time of the CW pulses was synchronized
with the electric pulses to the PM, so that only the photons in the
$\left\vert 1\right\rangle $ state were subject to the phase modulation. 
To provide accurate phase shifts, the electric pulse should be shorter than the difference
of the arrival time. The present circuit worked with the electric pulses of 10 ns duration. 
We used a high-speed LiNbO$_{3}$ PM designed for 10 Gb/s operation.
The CW and CCW pulses were combined by PBS1, and went back to the circulator.
Qubits were again represented by polarizations after PBS1.
PC2 compensated the birefringence but also provided Hadamard transformation. 
This was possible because PCs creates a fixed but arbitrary phase difference 
between two orthogonal polarizations, the axis of which can be set also at an arbitrary angle. 

\begin{figure} [pts]
\begin{center}
\includegraphics[
height=3.3295in,
width=4.7539in
]%
{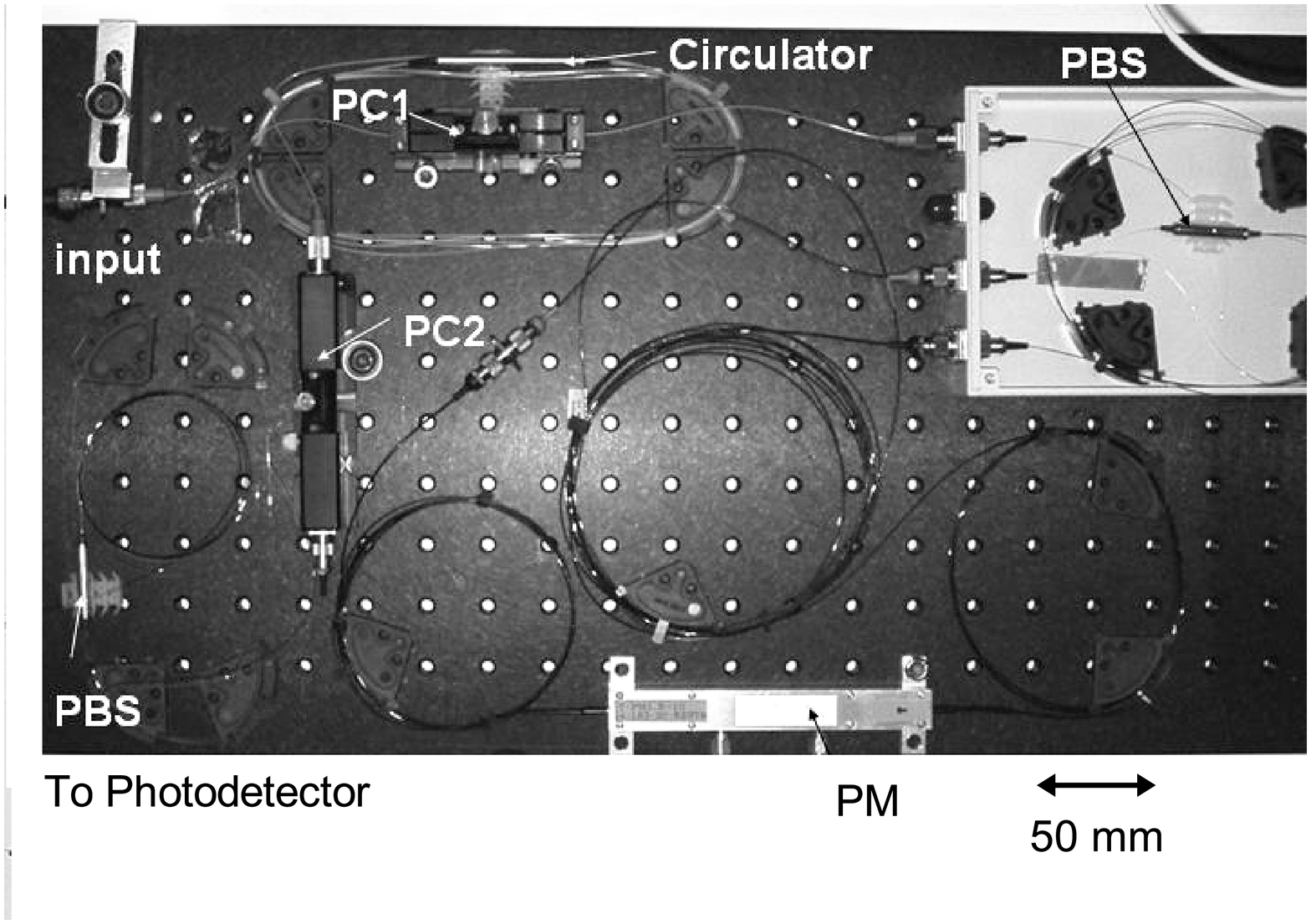}%
\caption{Quantum circuit for measured quantum Fourier transform constructed on a breadboard.} 
\label{figure3}%
\end{center}
\end{figure}

The polarizations of the output photons from the circulator were analyzed by PBS2.
Photons were detected by a photon detector.
The photon detector was made of balanced InGaAs avalanche photodiodes cooled
to 173 K to reduce the dark counts\cite{TN02}. We obtained the dark count rate 
at 6-7$\times$10$^{-7}$ per pulse with the
detection efficiency of 13 \%. The optical loss in the circuit (8.4
dB) came mainly from the insertion loss of the PM (6 dB). 
The whole circuit except the photon detector was constructed
on a breadboard of 350 mm$\times$600 mm, as shown in Fig. \ref{figure3}.

The input polarization state expressed by
\begin{equation}
\frac{1}{2^{1/2}}\left(  \left\vert 0\right\rangle +e^{2\pi i\,0.\varphi_{n}%
}\left\vert 1\right\rangle \right),  \cdots, \frac{1}{2^{1/2}}\left(
\left\vert 0\right\rangle +e^{2\pi i\,0.\varphi_{1}\cdot\cdot\cdot\varphi_{n}%
}\left\vert 1\right\rangle \right)  .\label{input-state}%
\end{equation}
was produced by a motorized polarization controller (Agilent 8169A) from the
pulses of a 1.55-$\mu$m laser diode. The pulse duration was about 30 ps. The light pulses were then
attenuated to 0.7 photons/pulse to ensure that the present circuit worked with a single photon.
The output bit values were compared with the input ones. 
They were also used to determine the phase modulation to the next qubit.

\begin{figure} [pts]
\begin{center}
\includegraphics[
height=2.47in,
width=3.98in
]%
{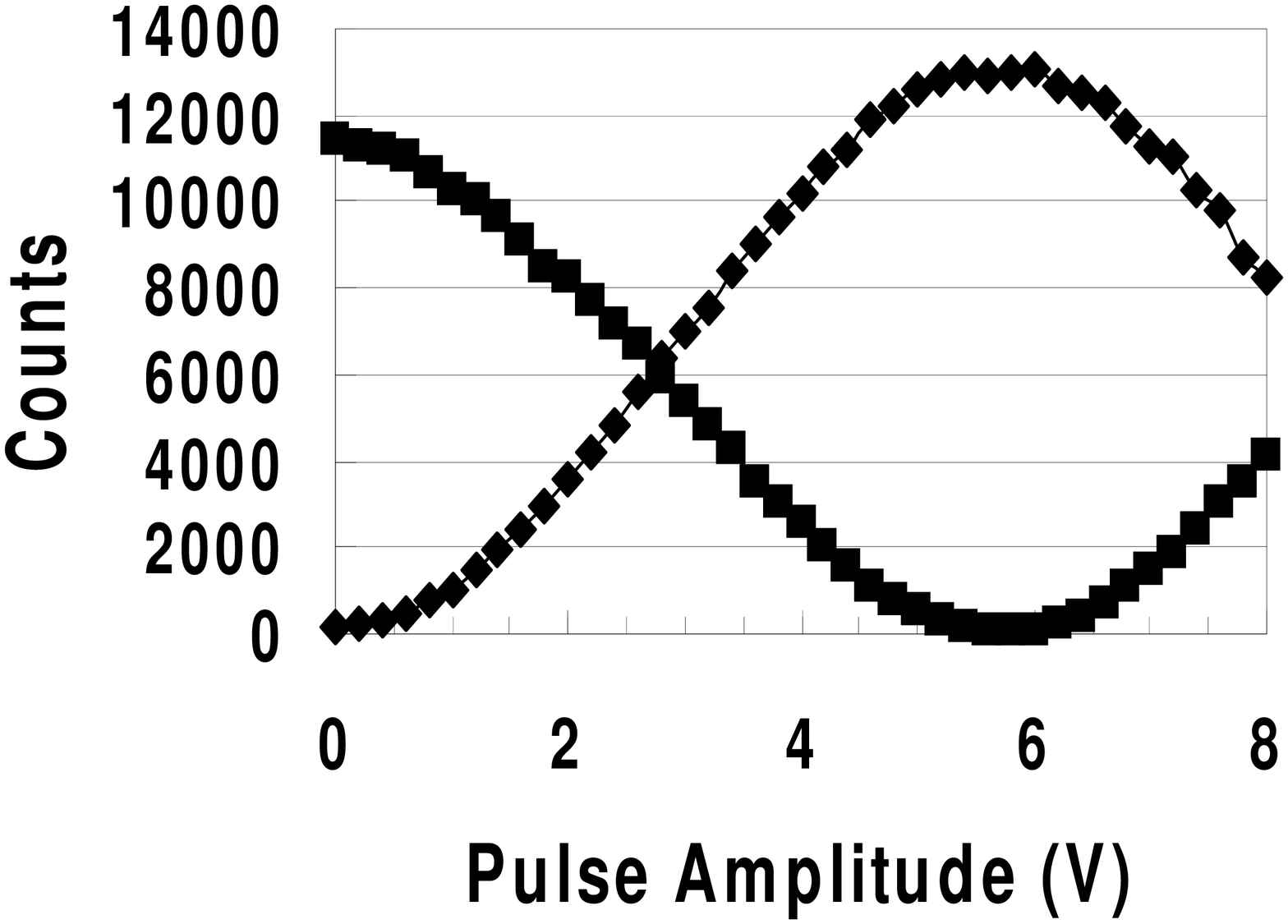}%
\caption{Interference fringe of the MQFT circuit. Polarization of the photons are set 
to 45 deg. The results were well fitted by cosine curves.}
\label{figure4}%
\end{center}
\end{figure}

\begin{figure} [pts]
\begin{center}
\includegraphics[
height=2.75in,
width=3.98in
]%
{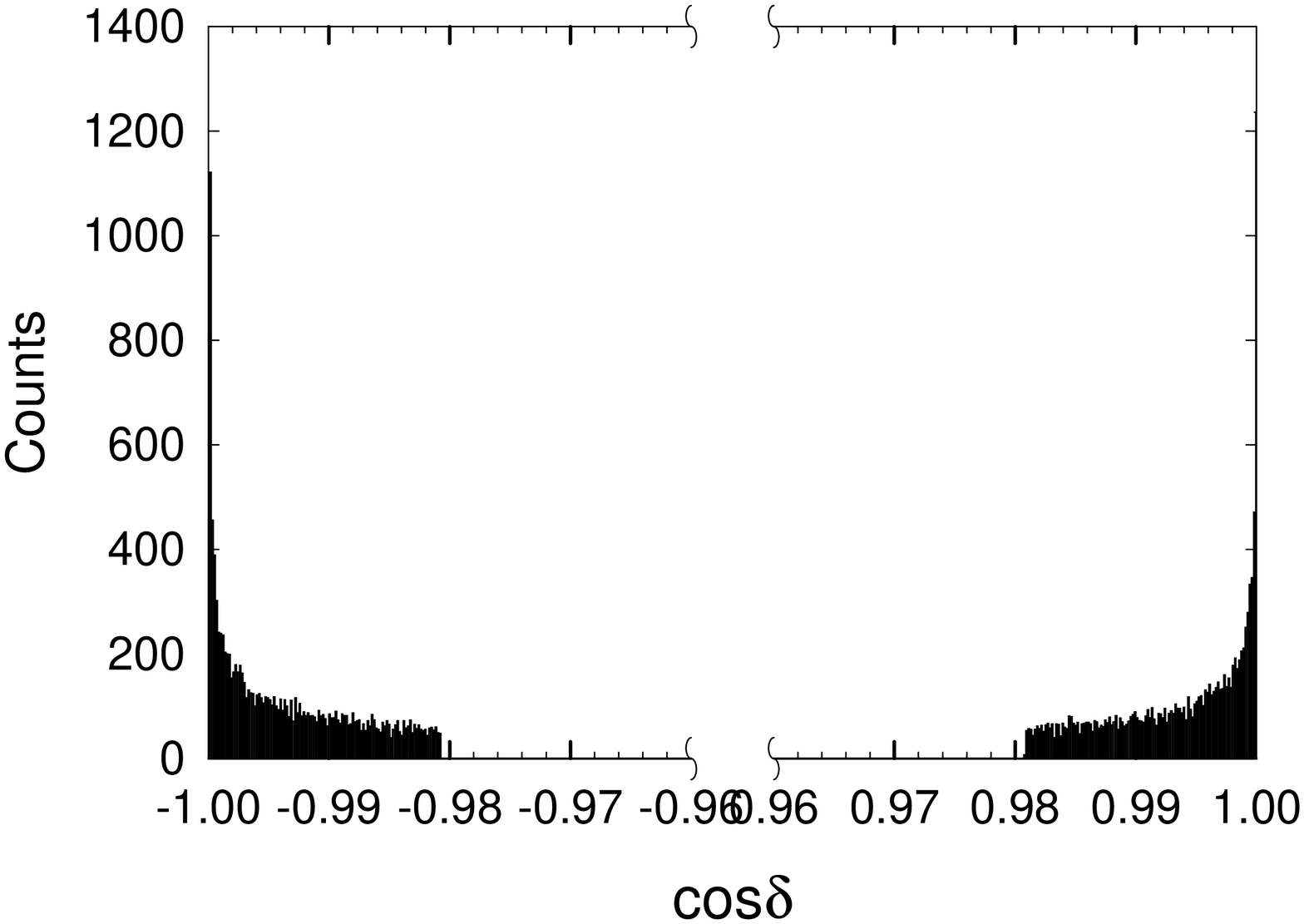}%
\caption{Distribution of the phase error in the phase modulation. The distribution was measured
in 24151 rotations for truncation $m$=5.}
\label{figure5}%
\end{center}
\end{figure}

Figure \ref{figure4} shows the output photons from the circuit as a function of the
drive voltage applied to the PM. The input polarization was fixed to the
linear polarization tilted at 45 deg. Visibility of the interferometer was
98 \%, which was determined mainly by the extinction ratio of the PBS. We
concluded that the phase modulation is linear to the drive pulse voltage,
because the interference signal was well fitted
to the cosine curve. 
The drive voltage for $\pi$ phase shift was 
estimated to be 5.80 V from Fig. \ref{figure4}. The phase errors measured in 24151 rotations for $m$=5
are shown in Fig. \ref{figure5}.  The values of $\cos\delta$ were distributed between
$[0.98,1]$ and $[-1,-0.98]$, with a mean value of $\left\langle \cos
\delta\right\rangle =\pm0.9936$.

\begin{figure} [pts]
\begin{center}
\includegraphics[
height=4.7539in,
width=3.3295in
]%
{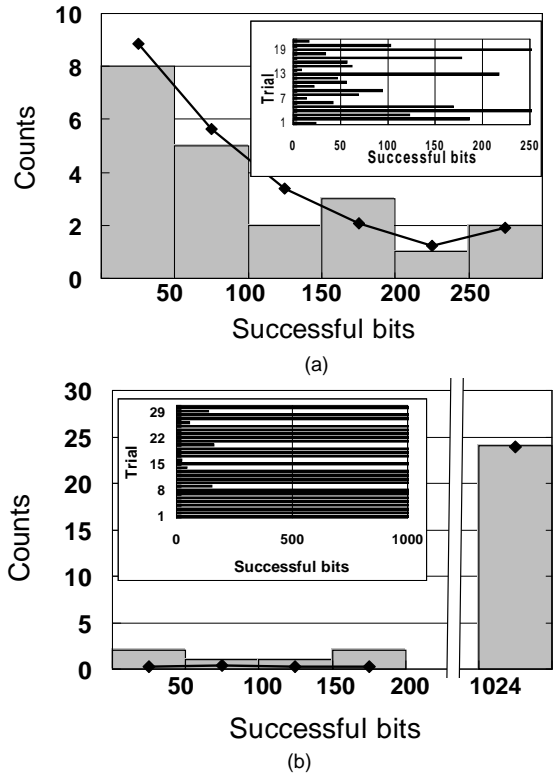}%
\caption{Results of MQFT trial, (a) 21 trials of 255 qubits and (b) 30 trials of 1024 qubits
The bit values in (b) were determined by the decision by majority of ten measurement outcomes.
The inset shows the
result of each trial. `Successfully transformed bits' in the figure refers to
the number of bits for which a MQFT operation was done successfully.}
\label{figure6}%
\end{center}
\end{figure}

Figure \ref{figure6}(a) shows the results of 21 trials of MQFT on 255 qubits. The inset shows the
result of each trial. `Successfully transformed bits' in the figure refers to
the number of bits for which a MQFT operation was done successfully. We
succeeded making a MQFT of full 255 qubits in two trials. The distribution of
the successful bits $n$ obeyed a geometric distribution with the error
probability per qubit $p$:
\begin{equation}
E(n)=(1-p)^{n-1}p.\label{geometric}%
\end{equation}
The average of the successfully transformed bits was 97, which related to the
error probability by $1/p$ in the geometric distribution. This reads the value
of the error probability $p=0.01$. More precisely, we estimated the range of
the error probability as follows. The probability of a
successful MQFTs up to the $k$th qubit n times in $N$ trials is expressed by%
\begin{equation}
P_{k}=\binom{N}{n}\left[  E\left(  k\right)  \right]  ^{n}\left[  1-E\left(
k\right)  \right]  ^{N-n}.\label{successprob}%
\end{equation}
Suppose the largest (smallest) number obtained in the experiment is $k_{max}$
($k_{min}$). The maximum value of the error probability per qubit $p_{max}$ is
defined so to yield a less than 5 \% probability of successful MQFTs for more
than $k_{max}$ qubits $N_{max}$ times. Similarly, the minimum value of the
error probability per qubit $p_{min}$ is defined so to yield a less than 5 \%
probability of successful MQFTs for less than $k_{min}$ qubits $N_{min}$
times. In the present experiment, we obtain $(k_{max},N_{max})=(255,3)$ and
$(k_{min},N_{min})=(9,1)$. The estimated error probability per qubit lay in
the range $1.2\times10^{-2}\leq p\leq$ $2.6\times10^{-3}$.

A simple way to reduce error probability is to decide the bit value by
majority of repeated measurements. We performed a MQFT of 1024 qubit to
examine the effect on the error reduction by the decision by majority. The
rotation and measurement were done with the same input polarization states
$M(=10)$ times, and the bit value was determined by the majority of the
results. We set the error probability per qubit at $p=0.07$, intentionally
higher than that for the previous measurement without accumulation. Figure
\ref{figure6}(b) shows the results of 30 trials of 1024 qubits. The success probability of
full 1024 qubits was 80\%, and the mean error probability was estimated to be
$2.2\times10^{-4}$ per qubit. More precisely, we obtained the range of the
error probability $1.2\times10^{-4}$ $\leq p\leq4.3\times10^{-4}$ with the
confidence level of 95 \% from the experimental results: $(k_{max}%
,N_{max})=(1024,24)$ and $(k_{min},N_{min})=(23,1)$. The error probability 
satisfies the error threshold for the fault-tolerant operation.

\section{Discussion}

\subsection{Effects of the imperfection} 

In the following, we consider effects of the imperfection of experimental apparatus.
Two types of failure may occur. One is failure in photon detection, the other is error in the
measurement. If a pulse contains no photons, the operation
will not affect the target qubits. We can thus continue the calculation by
repeating the operation step. If a photon is lost in the MQFT circuit, it
means that the c-$U$ and measurement have been done without knowing the
result. Therefore, the target states will be in a mixed state corresponding to
the two possible outcome. Even in this case, we can proceed the calculation by
repeating the same operation step. The repeated measurement will reduce the
target qubit state into a pure state by selecting one of the possibilities.

Errors in the measurement originate from imperfection of the interferometer
and from errors in the phase modulation. Dark counts are negligibly small in
the photon detector\cite{TN02}. The performance of the interferometer is
characterized by visibility $v$, which defines the measurement on the control
qubit as%
\begin{align}
M_{0} &  =\sqrt{\frac{1+v}{2}}\left\vert 0\right\rangle \left\langle
0\right\vert +\sqrt{\frac{1-v}{2}}\left\vert 1\right\rangle \left\langle
1\right\vert \nonumber\\
M_{1} &  =\sqrt{\frac{1-v}{2}}\left\vert 0\right\rangle \left\langle
0\right\vert +\sqrt{\frac{1+v}{2}}\left\vert 1\right\rangle \left\langle
1\right\vert .\label{measurement}%
\end{align}
The phase error, which shifts the rotation angle in (\ref{rotation}) from
$2\pi\Phi_{k}$ to $2\pi\Phi_{k}+\delta$, results in the control qubit after
the rotation gate as $2^{-1/2}\left(  \left\vert 0\right\rangle +\exp[2\pi
i\,0.\varphi_{n-k+1}^{s}+i\delta]\left\vert 1\right\rangle \right)  $. The
phase error originates from the approximation in the rotation angle $\Phi_{k}$
and from errors in converting the rotation angle into the drive voltage to the
PM. The latter can be reduced by careful calibration, so that we only have to
consider the effect of the truncation. The truncation at the $m$th bit results
in the phase error%
\begin{equation}
\delta=2\pi\sum_{j=m}^{k-1}\frac{1}{2^{j+1}}\varphi_{n-k+j+1}\leq2\pi
\sum_{j=m}^{k-1}\frac{1}{2^{j+1}}<2\pi\sum_{j=m}^{s\infty}\frac{1}{2^{j+1}%
}=\frac{\pi}{2^{m-1}}.\label{phaseerror}%
\end{equation}
The phase error should not be significant\cite{BEST96}, because the
contribution from the $j$th bit ($j>m$) decreases with the factor of
$2^{-(j+1)}$. The worst values of $\cos\delta=\pm0.98$ obtained in the
experiment corresponds to a phase error of $\pi/16$, which agrees quite well
with the prediction by (\ref{phaseerror}) with $m$=5. The visibility and
the phase error determine the error probability of the measurement by
\begin{equation}
p=\frac{1-v\cos\delta}{2}.\label{error}%
\end{equation}
The estimated error probabilities from (\ref{error}), $p=8.2\times10^{-3}$
(by using $\left\langle \cos\delta\right\rangle =\pm0.9936$) and
$p=1.5\times10^{-2}$ (by using $\cos\delta_{\max}=\pm0.98$), agree well with
the experiment.

\subsection{Validity of majority voting}

We consider the validity of decision by majority of repeated measurements. If
the state of the target qubits is the eigenstates of the c-$U$, the unitary
transform results in the same phase value to the control qubit. However, in
most case, the target qubit state is a superposition of the eigenstates. In
this case, the total state at the $k$ th operation step after the rotation
gate is transformed by the Hadamard gate to%
\begin{equation}
\left\vert \Psi\right\rangle =\left(  \frac{1+e^{i\delta}}{2}\left\vert
0\right\rangle +\frac{1-e^{i\delta}}{2}\left\vert 1\right\rangle \right)
\left\vert u_{0}^{s}\right\rangle +\left(  \frac{1-e^{i\delta}}{2}\left\vert
0\right\rangle +\frac{1+e^{i\delta}}{2}\left\vert 1\right\rangle \right)
\left\vert u_{1}^{s}\right\rangle ,\label{errorstate}%
\end{equation}
where the states $\left\vert u_{0}^{s}\right\rangle $ and $\left\vert
u_{1}^{s}\right\rangle $ defined by
\begin{align}
\left\vert u_{0}^{s}\right\rangle  &  =C_{0}\sum_{s\in\left\{  \varphi
_{n-k+j+1}^{s}=0\right\}  }c_{s}\left\vert u_{s}\right\rangle \nonumber\\
\left\vert u_{1}^{s}\right\rangle  &  =C_{1}\sum_{s\in\left\{  \varphi
_{n-k+j+1}^{s}=1\right\}  }c_{s}\left\vert u_{s}\right\rangle
\label{branchstates}%
\end{align}
are normalized as to $\left\langle u_{0}^{s}|u_{0}^{s}\right\rangle
+\left\langle u_{1}^{s}|u_{1}^{s}\right\rangle =1.$ Suppose measurement
outcome is "0", then the state collapse to
\begin{equation}
\frac{1}{\sqrt{P_{0}}}M_{0}\left\vert \Psi\right\rangle =\frac{1}{\sqrt{P_{0}%
}}\left[  \sqrt{\frac{1+v}{2}}\left\vert 0\right\rangle \left\vert \tilde
{u}_{0}\right\rangle +\sqrt{\frac{1-v}{2}}\left\vert 1\right\rangle \left\vert
\tilde{u}_{1}\right\rangle \right]  ,\label{collapse}%
\end{equation}
where $P_{0}=\left[  1+\left(  2\left\langle u_{0}^{s}|u_{0}^{s}\right\rangle
-1\right)  v\cos\delta\right]  /2$ is the probability to obtain the outcome
"0", and the states $\left\vert \tilde{u}_{0}\right\rangle $ and $\left\vert
\tilde{u}_{1}\right\rangle $ are defined by
\begin{align}
\left\vert \tilde{u}_{0}\right\rangle  &  =\frac{1+e^{i\delta}}{2}\left\vert
u_{0}^{s}\right\rangle +\frac{1-e^{i\delta}}{2}\left\vert u_{1}^{s}%
\right\rangle \nonumber\\
\left\vert \tilde{u}_{1}\right\rangle  &  =\frac{1-e^{i\delta}}{2}\left\vert
u_{0}^{s}\right\rangle +\frac{1+e^{i\delta}}{2}\left\vert u_{1}^{s}%
\right\rangle .\label{newbasis}%
\end{align}
Partial trace on the control qubit yields a density matrix of the target qubit
states%
\begin{equation}
\rho=\frac{1}{P_{0}}\left[  \frac{1+v}{2}\left\vert \tilde{u}_{0}\right\rangle
\left\langle \tilde{u}_{0}\right\vert +\frac{1-v}{2}\left\vert \tilde{u}%
_{1}\right\rangle \left\langle \tilde{u}_{1}\right\vert \right]
\label{densitymatrix}%
\end{equation}
The c-$U$ and Hadamard transform (\ref{errorstate}) followed by the measurement
(\ref{measurement}) on the density matrix (\ref{densitymatrix}) results in the
same value of the error probability (\ref{error}). Therefore, the accumulation
of $M$ measurement outcomes will reduce the error probability to%
\begin{equation}
p_{M}=\sum_{j=0}^{\left[  M/2\right]  }\binom{M}{j}p^{M-j}\left(  1-p\right)
^{j},\label{reduced}%
\end{equation}
which decreases rapidly for $p\ll1$. For example, the error probability per
operation $p=0.07$ and accumulation $M=10$ yield the error probability
$p_{10}=\allowbreak3\times10^{-4}$, which agrees with the value obtained from
the experiment on 1024 qubits.

\subsection{Toward quantum computers}

The main drawback of the MQFT is time, because it operates serially. 
The response time of the current circuit is limited by electronics. 
We may expect the response time to be reduced to several nanoseconds. 
If the operation time is decreased to 1 ns, the target qubits should 
remain coherent for at least 10 $\mu$s to complete controlled-unitary operations 
with a thousand control qubits (1 ns $\times$ 1000 qubits $\times$ 10 accumulations.) 
This coherence time would be possible, because the coherence time between 
the meta-stable states of an atom reaches 0.5 ms\cite{Ph01}. 
Here, we consider a quantum computer consists of photon control qubit and 
atom target qubits. The input photons induce a unitary transformation on the 
atom qubits. The controlled operation can be achieved by the polarization selection
rules in the atomic transitions, for example. The photons are scattered by the atoms
and obtain a phase shift through the kick-back effect. The phase shift will be analyzed 
by MQFT and provide a solution of the problem. 
This scheme shows an interesting analogy between the quantum computation and the spectroscopy;
the change in the system (target qubits) state is probed by the change 
in the scattered photon state (control qubit.)

The remaining
problems are in realizing controlled-unitary operations; the QFT by itself
will not provide an exponential speed up in comparison with classical
algorithms. 
One possibility is using atomic target qubits as discussed above, where the unitary transformation
is achieved by the interaction between atoms. We need to construct unitary transformations in the atomic qubits,
so that it may be seen just moving the difficulty to the atoms. 
Nevertheless, we believe this scheme would make the problem simpler.
It can exploit fairly large interaction between atomic qubits to obtain two qubit gates.
Measurement on photonic control qubit will resolve the read-out problem.
Another possibility is to combine with the linear-optics gates proposed by Knill, Laflamme, and Milburn\cite{KLM01}.
Since the KLM scheme utilizes single photon states, it suits well to the present MQFT circuit.

The present MQFT circuit may find its applications on
such as finding eigenvalues and eigenvectors\cite{AL99} and
clock synchronization\cite{Chuang00} in the near future. In these
applications, fewer controlled-unitary operations are enough to achive a meaningful task than those
required in the factorization algorithm.

\section{CONCLUSION}

We have shown successful MQFT using single photon by the use of
commercially available fiber optic devices. The MQFT is robust to
errors as suggested. We have also shown that decision by majority is simple to
implement and is useful to reduce error probability. Our implementation can be
used in the final part of the order-finding algorithm. In other words, the
'easier part' of the Factorization algorithm is ready. The present MQFT
circuit decreases the qubit-gate-operation product, and thus reduces the
requirement for the threshold on the error probability. We believe the present
MQFT implementation provides a useful base to realize a large scale quantum computation.

\section*{Acknowledgements}

The authors would like to thank Dr. S. Ishizaka, Prof. K. Matsumoto, Dr. M. Hayashi, Dr. M.
Iinuma, and Prof. S. Takeuchi for fruitful discussions.

\end{document}